\documentclass{elsarticle}
\usepackage{epsfig}
\journal{Physics Letters B}

\newcommand{\be}{\begin{equation}}
\newcommand{\ee}{\end{equation}}

\begin{document}

\begin{frontmatter}

\title{The vector meson mass in the large $N$ limit of QCD}
\author{A. Hietanen, R. Narayanan, R. Patel and C. Prays}
\address{
Department of Physics, Florida International University, Miami,
FL 33199.}

\begin{abstract}
The vector meson mass is computed as a function of quark mass
in the large $N$ limit of QCD. We use continuum reduction and
directly compute the vector meson propagator in momentum
space. Quark momentum is inserted
using the quenched momentum prescription. 
\end{abstract}

\begin{keyword}

Large $N$ QCD \sep Vector meson masses \sep Low energy constants

\end{keyword}

\end{frontmatter}

Meson masses remain finite in the 
't Hooft limit of large $N$ QCD
in four dimensions~\cite{Manohar:1998xv}.
Chiral symmetry is broken and the value of
the chiral condensate has been measured on the lattice
for overlap fermions
using random matrix theory techniques~\cite{Narayanan:2004cp}.
The result can be summarized as~\cite{Narayanan:2005gh}
\be
\frac{\Sigma(b)}{T_c^3(b)} = 0.828 \left[ \ln \frac{0.268}{T_c(b)}
\right]^{\frac{9}{22}}
\label{cond}
\ee
where $b=\frac{1}{g^2N}$ is the bare
't Hooft coupling on the lattice. 
The deconfining temperature, $T_c(b)$, is also 
known from a lattice calculation~\cite{Kiskis:2003rd} and
is given by
\be
b_I = b e(b);~~~~~~~e(b)=\frac{1}{N} 
\langle {\rm Tr} U_p (x) \rangle;~~~~~
T_c(b)=3.85 \left ( \frac {48\pi^2 b_I }{11}\right )^{\frac{51}{121}} 
e^{-\frac{24\pi^2 b_I}{11}}.
\ee

Continuum reduction holds 
if $L> \frac{1}{T_c(b)}$~\cite{Kiskis:2003rd} and meson
propagators can be directly computed in Euclidean momentum space
without any finite volume effects. The pion mass as a function
of quark mass, $m_o$, 
was computed on the lattice using overlap fermions
and the pion decay constant is given by~\cite{Narayanan:2005gh} 
\be
\frac{f_\pi}{\sqrt{N} T_c(b)} = 0.269. \label{fpi} 
\ee

In this letter, we present results for
the mass of the vector meson, $m_\rho$,
as a function of the quark mass, $m_o$, using the same technique
as the one used for the computation of the pion mass 
in~\cite{Narayanan:2005gh}. The $\rho$ propagator is computed
using 
\be
{\cal M}_{\mu\nu} (p,m_o) = {\rm Tr} \left [
S \gamma_\mu G(U_\mu e^{\frac{ip_\mu}{2}},m_o) 
S \gamma_\nu G(U_\mu e^{-\frac{ip_\mu}{2}},m_o)\right ].
\ee
\begin{itemize}
\item $G(U_\mu, m_o)$ is the lattice quark propagator computed
using overlap fermions in a gauge field background given by
$U_\mu$.
\item The phase factors,  $e^{\pm\frac{ip_\mu}{2}}$, multiplying
the gauge fields correspond to the force-fed momentum of the
two quarks in the quenched momentum prescription.
\item The meson momentum was chosen to be
\be
p_\mu=\cases{ 0 & if $\mu =1,2,3$ \cr \frac{2\pi k}{NL};\ \ \
k=2,3,4,5,6 & if $\mu=4$.}\label{momen}
\ee
\item $S$ smears the operator in the zero momentum directions
using the inverse of the gauged Laplacian.
\end{itemize}
The $\rho$ meson is made up of two different quarks (say $u$ and $d$)
with degenerate quark masses. Since the associated vector currents
are conserved, the propagator, after averaging over gauge
fields, will be of the form
\be
{\cal M}_{\mu\nu} (p,m_o) = \frac{ A
\left (p_\mu p_\nu - p^2\delta{\mu\nu}\right)}{p^2 + m_\rho^2(m_0)},
\label{rhoprop}
\ee
assuming the propagator is
dominated by the lowest vector meson state.
Our numerical result is consistent with the above form.
We found all off-diagonal ($\mu\ne\nu$) terms and the $\mu=\nu=4$
term to be zero within errors for the specific choice of momentum
in (\ref{momen}) and we also found the $\mu=\nu=1,2,3$ terms to be
the same within errors in our small test runs.
Since the evaluation of the quark propagators
is the computationally intensive part,
we set $\mu=\nu=1$ and obtained a value for the $\rho$ meson mass
at six different quark masses by fitting it to the form 
in (\ref{rhoprop}).

\begin{figure}
\epsfxsize = 0.8\textwidth
\centerline{\epsfbox{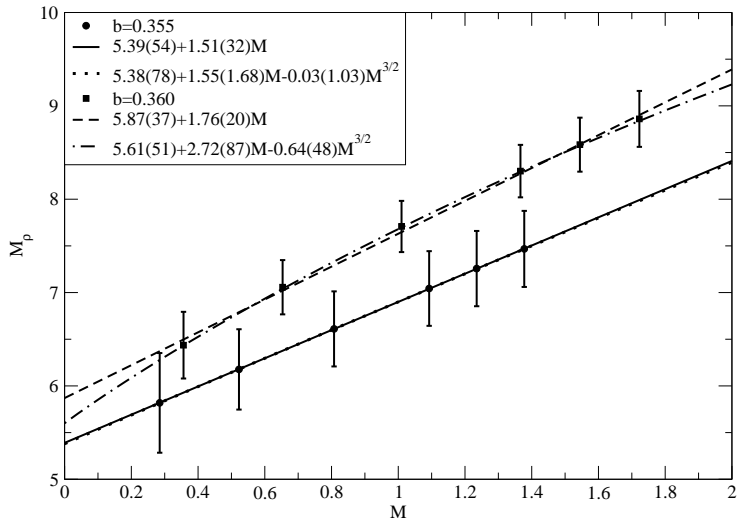}}
\caption{A plot of $\rho$ mass as a function of the renormalization group invariant quark mass in dimensionless units.}
\label{rho}
\end{figure}

Four different couplings were used in~\cite{Narayanan:2005gh}
for the computation of the meson masses. We found two of those
couplings to be too strong for the computation of the $\rho$ mass.
We report here, the results for the
$\rho$ mass at two couplings, namely,
$b=0.355$ and $b=0.360$. Chiral perturbation theory for
vector mesons~\cite{Jenkins:1995vb} suggests that we fit the
data to the form
\be
M_\rho = \bar{\cal M}_8 + \Lambda_2 M + \delta M_\rho,
\label{mrho}
\ee
where
\be
M_\rho = \frac{m_\rho}{T_c(b)};
\ \ \ 
M = 2 \frac{m_0 \Sigma}{ T_c^4(b)};\ \ \ 
\ee
are the mass of the $\rho$ meson and
the renormalization group invariant quark mass\footnote{
$M$ denoted the sum of the two quark masses comprising the
$\rho$ meson and hence the factor of
$2$ in the formula for $M$.}
measured in units of the deconfining temperature.
The two coefficients in (\ref{mrho}) are two of the coefficients
in the mass term in the chiral lagrangian, namely,
\be
\bar{\cal M}_8 = \frac{\bar\mu_8}{T_c(b)};\ \ \ \
\Lambda_2 =  \frac{\lambda_2 T_c^3(b)}{\Sigma}.
\ee

The data is plotted in Fig.~\ref{rho}. The result for
the chiral condensate in (\ref{cond}) was obtained such
that the pion mass as a function of the quark mass 
scaled properly in the range of coupling from $b=0.345$
to $b=0.360$. It is not necessary that this eliminates
finite lattice spacing effects on all quantities and we
do see effects of finite lattice spacing
effects in Fig.~\ref{rho}.
A linear fit performs quite well
at both the couplings to yield consistent estimates for
$\bar{\cal M}_8$ and $\Lambda_2$. The results are
shown both in Fig.~\ref{rho}
and Table~\ref{tab1}. 

\begin{table}
\centerline{
\begin{tabular}{|c|c|c|c|c|c|c|}
\hline
&&&&&&\\
$L$ & $N$ & $b$ & $T_c(b)$ & $\Sigma^{1/3}(b)$ & $\bar{\cal M}_8$ & 
$\Lambda_2$ \\
&&&&&&\\
\hline
&&&&&&\\
10 & 19 & 0.355 & 0.144 & 0.1265 & 5.39(54) & 1.51(32) \\
11 & 17 & 0.360 & 0.125 & 0.1130 & 5.87(37) & 1.76(20) \\
&&&&&&\\
\hline
\end{tabular}}
\caption{ Simulation parameters, critical box size, bare chiral
condensate along with the estimates for the two
coefficients $\bar{\cal M}_8$ and $\Lambda_2$ in the
mass term of the chiral lagrangian.}
\label{tab1}
\end{table}

Chiral perturbation theory suggests that $\delta M_\rho$
in (\ref{mrho}) should lead off as $M^{\frac{3}{2}}$ and
the coefficient of this leading term should be negative.
A fit with a $M^{\frac{3}{2}}$ term is shown in Fig.~\ref{rho}
and we see that the coefficient at $b=0.360$ is consistent
with it being negative. The error in this coefficient
is rather large.

Using the result for $\bar{\cal M}_8$ in Table~\ref{tab1} for
$b=0.360$ and the result for $f_\pi$ in (\ref{fpi}), we have
\be
\bar\mu_8 = \frac{21.8 \pm 1.4}{\sqrt{N}} f_\pi.
\ee
If we use $f_\pi=86$ MeV and $N=3$, then we get 
$\bar\mu_8=1082 \pm 70$ MeV.

The vector meson masses have been computed in the quenched
approximation for $N=2,3,4,6$ in~\cite{DelDebbio:2007wk,Bali:2008an}.
The couplings used in~\cite{DelDebbio:2007wk} and in~\cite{Bali:2008an}
are roughly the same.
The strongest and weakest coupling correspond to 
$b=0.296$ and $b=0.353$ respectively in the notation of this paper. 
There is a bulk transition on the lattice in the large $N$ limit
that becomes a cross-over at finite $N$. The region between
$b=0.34$ and $b=0.36$ is in the meta-stable region of this
transition~\cite{Kiskis:2003rd} 
and we need to be above $b=0.34$ to be in the
continuum phase of the large $N$ theory. Since the vector meson
is heavy compared to the pion for small quark masses, finite
lattice spacing effects are larger in the case of the vector meson.
Our study at $b=0.350$,
not reported in this paper, does yield a value for $\bar{\cal M}_8$
that is about $25\%$ smaller than the one quoted here at $b=0.360$
and consistent with the value obtained in~\cite{Bali:2008an}.

\section*{Acknowledgments}

A.H. and R.N. 
acknowledge partial support by the NSF under grant number
PHY-055375. 
A.H. also acknowledges partial support by US Department
of Energy grant under contract DE-FG02-01ER41172.
R.N. would like to thank Joe Kiskis for some useful
discussions.

\end{document}